\documentclass[aps,physrev,reprint,superscriptaddress]{revtex4-2}

\usepackage{graphicx}
\usepackage{dcolumn}
\usepackage{bm}
\usepackage[separate-uncertainty=true]{siunitx}

\begin{document}


\title{Real-Time Cryo-CMOS Photon-Number Readout from a Single SNSPD}

\author{Niklas Lamberty}
\email[corresponding author: ]{niklas.lamberty@uni-paderborn.de}
\affiliation{Department of Physics, Paderborn University, Warburger Str. 100, 33098, Paderborn, Germany}
\affiliation{Institute for Photonic Quantum Systems (PhoQS), Paderborn University, Warburger Str. 100, 33098, Paderborn, Germany}
\author{Dileep V. Reddy}
\affiliation{National Institute of Standards and Technology, Boulder, Colorado 80305, USA}
\affiliation{Department of Physics, University of Colorado, Boulder, Colorado 80309, USA}
\author{Aishi Barua}
\affiliation{Department of Physics, Paderborn University, Warburger Str. 100, 33098, Paderborn, Germany}
\affiliation{Institute for Photonic Quantum Systems (PhoQS), Paderborn University, Warburger Str. 100, 33098, Paderborn, Germany}
\author{Martin J. Stevens}
\affiliation{National Institute of Standards and Technology, Boulder, Colorado 80305, USA}
\author{Tim J. Bartley}
\affiliation{Department of Physics, Paderborn University, Warburger Str. 100, 33098, Paderborn, Germany}
\affiliation{Institute for Photonic Quantum Systems (PhoQS), Paderborn University, Warburger Str. 100, 33098, Paderborn, Germany}

\date{\today}

\begin{abstract}
Measurements leveraging the intrinsic photon-number resolution of Superconducting Nanostrip Single-Photon Detectors (SNSPDs) have recently gained traction due to the  simplicity of the detector design. Intrinsic photon-number measurement schemes using SNSPDs encode the photon-number information in picosecond signal timing; these systems thus require high bandwidth readout and sensitive electrical measurement equipment. We show a cryogenic electronic circuit operating close to the SNSPD in the cryostat, which can isolate single-photon detection events directly from the SNSPD signal. This reduces the latency of the photon-number readout to \SI{13.9\pm0.3}{ns}, which enables feedforward operations based on intrinsic photon-number measurements with low latency.
\end{abstract}


\maketitle

The accurate measurement of photon numbers is required for multiple quantum optical applications \cite{Hadfield2009}. Such applications include Gaussian Boson sampling \cite{Hamilton2017,Stefszky2026,Fu2026}, loss detection in gates \cite{Lim2005,DeGliniasty2024,Chan2026} or fusion \cite{Browne2005,Bartolucci2023,Hauser2025} in universal optical quantum computing and the heralding of Fock states \cite{Tiedau2019,Sonoyama2024}. A major challenge in achieving high fidelity Photon-Number Resolved (PNR) measurements is photon loss, which reduces the number of photons registered by the detector and thereby causes measurement errors. Superconducting Nanostrip Single Photon Detectors (SNSPDs) have therefore seen widespread adoption for this purpose as they can achieve detection efficiencies of up to \SI{98}{\%} \cite{Reddy2020}.

The primary challenge in using SNSPDs for PNR measurements is that these detectors are not inherently energy resolving (like Transition Edge Sensors (TES) \cite{Cabrera1998,Lita2008})
and thus require additional optics or advanced readout schemes to achieve PNR. Common techniques for achieving PNR are the splitting of the measured state onto multiple SNSPDs \cite{Natarajan2013,Hauser2025} or using multi-pixel SNSPD arrays \cite{Dauler2007,Tiedau2020,Resta2023}. However, both of these methods suffer from diminishing multi-photon efficiencies as all photons have to impinge on different detection elements to be detected \cite{Stasi2023}. 

Recently a new technique for PNR measurements with single element SNSPDs has been developed, which relies on counting the number of resistive hotspots created by simultaneous photon detection events on a single SNSPD \cite{Cahall2017,Nicolich2019}. The number of hotspots is directly related to the slew rate of the rising edge of the electrical signal generated by the SNSPD. Since the change in slope is typically in the range of tens of picoseconds, high precision Time-to-Digital Converters (TDC) \cite{Sauer2023,Los2024,Schapeler2024} or oscilloscopes \cite{Endo2021,Schapeler2024} are required to extract the photon-number information. This approach, however, faces bottlenecks both in scalability as well as in processing latency.

For applications utilizing feedforward operation like photon source multiplexing \cite{Kaneda2019} and certain quantum gates \cite{DeGliniasty2024}, operations have to be performed based on a photon-number measurement result. Since long term storage of quantum states is difficult, feedforward operations have to be performed with a low latency. However, achieving such low latency is challenging when using TDCs and oscilloscopes as they are complex general purpose analysis tools with significant internal processing latencies. Room temperature electronics are additionally latency limited by long cables to the cryogenic detectors, motivating the use of cryogenic electronic circuits which are mounted close to the detectors\cite{Thiele2025}. 

Photon-number readout using external processing is also limited in scaling as every detector requires a high-bandwidth RF-cable to room-temperature as well as a dedicated channel on the TDC/oscilloscope. By performing the photon-number readout at the cryogenic stage the resulting digital signal can be transferred through cables with lower bandwidth and noise requirements and analyzed by a simple counter. Digital readout of the photon-number also makes it possible to serialize the readout signal and thereby use a single cable to readout multiple detectors \cite{Takeuchi2020}.

\begin{figure}[ht]
    \centering
    \includegraphics[width=1\linewidth]{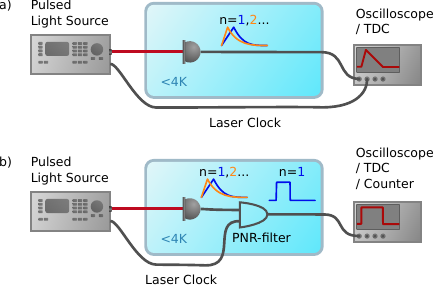}
    \caption{a) The conventional method for reading out photon-number information from a single SNSPD. The analog detection signal of the SNSPD is transmitted to room temperature and analyzed by an oscilloscope or Time-to-Digital Converter (TDC). The signal for the single photon (n=1) detection case is shown in blue, and for the two photon (n=2) selection case in orange. b) The implemented cryogenic photon-number readout. The analog detection signal is analyzed on the \SI{4}{K} stage of the cryostat and a digital signal signifying a single photon detection is transmitted to room temperature.}
    \label{fig:abstract}
\end{figure}

In this paper we present a cryogenic electronic circuit which can select photon numbers from the electrical signal generated by a high-efficiency SNSPD. The circuit then generates a digital signal when a desired photon-number has been detected, as shown in Fig. \ref{fig:abstract} for the single-photon selection case. We characterize the photon-number selection capabilities of the circuit and apply it as a real-time single photon herald for a parametric down-conversion source. 

\section{Methods}

Photon-number selection is achieved by comparing the time difference between a clock (derived from the pulsed laser source) and the rising edge of the SNSPD detection signal \cite{Sauer2023,Los2024}. In this measurement, a shorter delay between the laser and the SNSPD signal corresponds to a steeper rising edge and therefore a higher number of detected photons as shown in Fig.~\ref{fig:abstract}. In order to achieve photon-number selectivity the circuit must therefore pass only detection events with a large delay to set a photon-number cutoff. The circuit thereby acts as a low-pass filter for photon-numbers and thereby enables photon-number selectivity. 

\begin{figure}
    \centering
    \includegraphics[width=\linewidth]{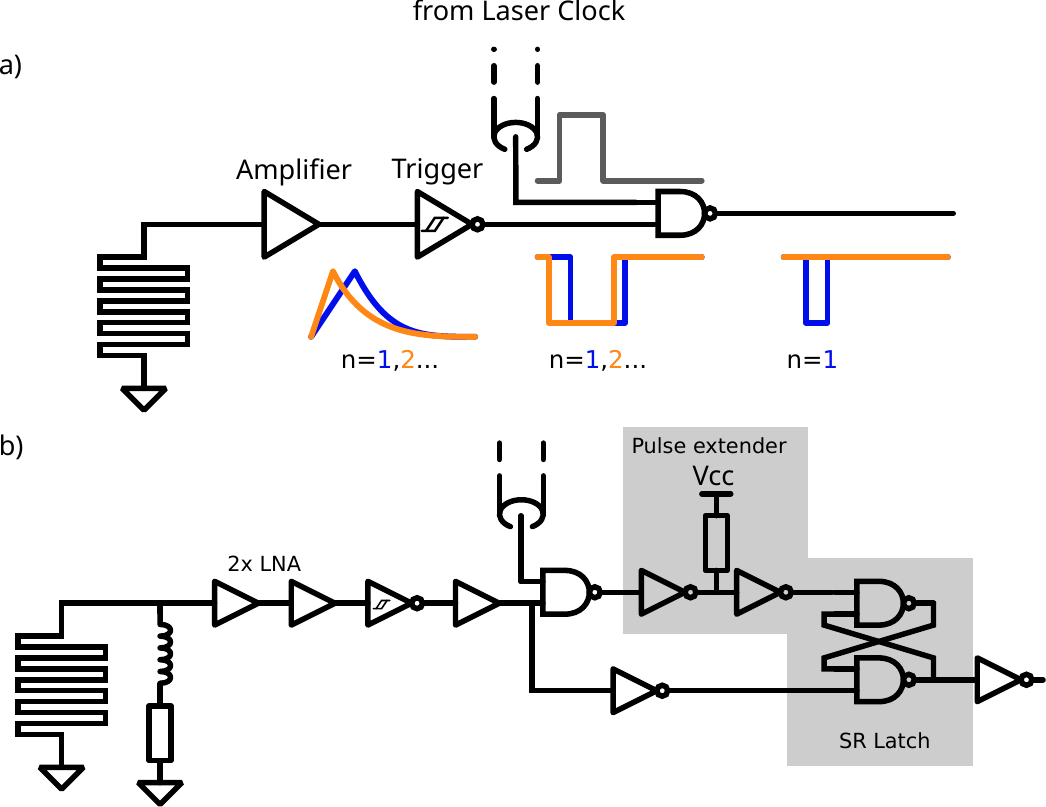}
    \caption{a) A simplified circuit diagram of the photon-number selection circuit set to pass only single-photon events. The voltage traces next to the wires illustrate the signal at the different stages of the circuit. The traces show a single photon detection event (blue), a two photon detection event (orange) and the external clock signal (gray). b)~The complete schematic of the photon-number selection circuit. The additional logic cleans the output pulse and the inductive shunt prevents the SNSPD from latching.}
    \label{fig:shematic}
\end{figure}

The working principle of the circuit is illustrated in Fig.~\ref{fig:shematic}~(a). The circuit first transforms the SNSPD detection signal from a millivolt-level analog signal into a \SI{3}{V} digital signal for the subsequent timing analysis. The signal is then overlapped with an external signal derived from the laser clock on a NAND gate. By correctly adjusting the timing overlap, higher-order photon-number detection events can be filtered out.

In addition to the core parts of the photon-number discrimination circuit seen in Fig.~\ref{fig:shematic}~(a) we include elements in the experimental setup to simplify the readout and operation of the system, as can be seen in Fig.~\ref{fig:shematic}~(b). First we add a set-reset (SR) latch at the output of the NAND gate. This latch is set by the single-photon detection signal and reset with the inverted SNSPD signal. This latch thereby generates a digital output signal of a length similar to the SNSPD signal, in the order of \SI{100}{ns}, instead of the short single-photon detection signal, in the order of \SI{100}{ps}, generated at the NAND gate. This SR-latch has the additional function of blocking signal generation in the absence of an SNSPD detection signal, as the reset of the SR-latch stays active when the system receives no SNSPD signal and thereby prevents an output from being generated.

Another addition to the photon selection circuit is a pulse extender to stretch the single-photon detection signal from the NAND gate. This is required due to the setup time of the SR-latch, which starts to oscillate if the set input receives a too short pulse. We additionally add an inductive shunt for the SNSPD to prevent latching of the SNSPD\cite{DeCea2020}. 

To relax the timing requirements for the photon-number selection circuit, we utilize a high-inductance SNSPD. This increases the rise time of the SNSPD thereby making the temporal separation between the photon numbers larger \cite{Kong2024}. This additionally has the benefit of providing a high detection efficiency of \SI{86\pm5}{\%} due to the large area of the detector.

The whole circuit is assembled from individually packaged CMOS logic gates, which are soldered on a printed circuit board. The trigger used to digitize the SNSPD signal is the same as in ref.\cite{Thiele2025}. All electronics are mounted on the \SI{4}{K} stage of the cryostat while the SNSPD is mounted to the \SI{1}{K} stage.

\section{System Characterization}\label{ch:characterisation}

\begin{figure*}
    \centering
    \includegraphics[width=0.9\linewidth]{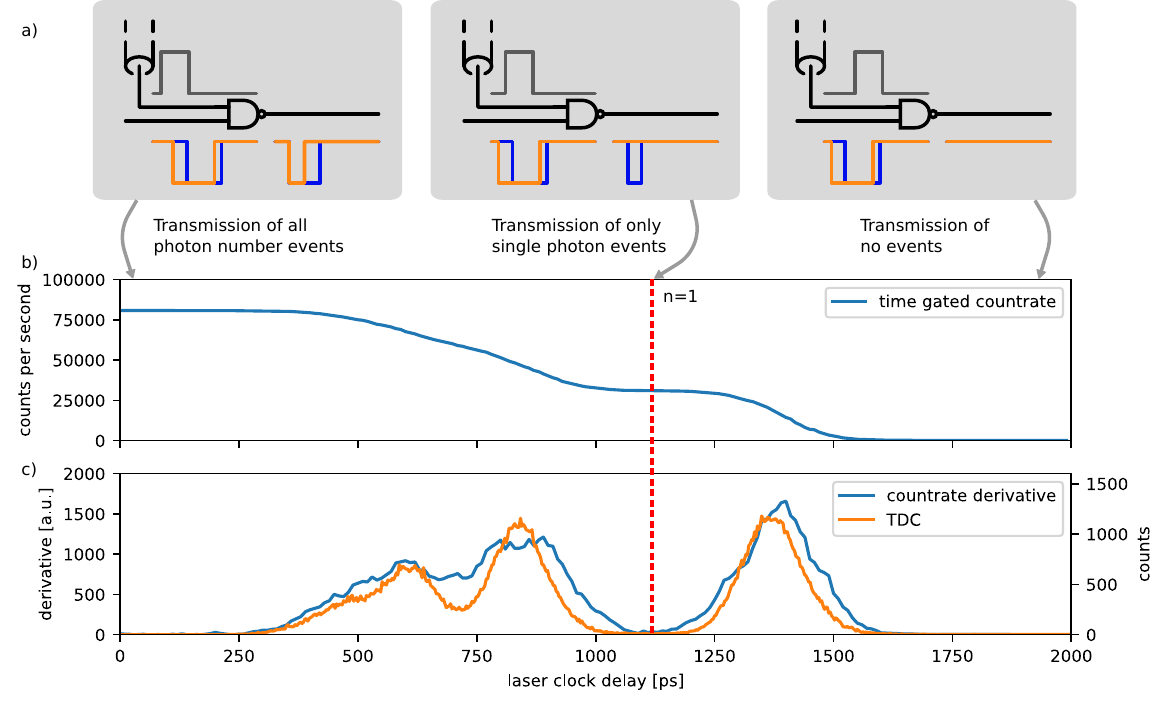}
    \caption{a) The different operating regimes of the photon-number filtering circuit depending on the delay of the laser clock, where either all photon numbers, only single photons or no events are passed through the NAND gate. b) The measured count rate after the photon-number selection circuit measured against the delay of the external laser clock. The plateau region at around \SI{1100}{ps} indicates the circuit selecting only single photon detection events. The red line indicates the optimal delay for operating as a single photon filter. c) The derivative of the count rate shown in b) as well as the rising edge delay distribution of the SNSPD measured with a Time-to-Digital Converter (TDC). The two curves show good overlap.}
    \label{fig:coherent_state_characterisation}
\end{figure*}

We first characterize the function of the photon-number selection circuit by illuminating the SNSPD with Poisson distributed photon numbers. To do so, we use a pulsed laser diode with a wavelength of \SI{1550}{nm} and \SI{9}{ps} pulse duration, which is attenuated to the single photon level. By gradually shifting the delay of the laser clock relative to the SNSPD signal, we observe the gradual rejection of faster rising edges, as shown in Fig.~\ref{fig:coherent_state_characterisation}~(a,b).

For better visualization, the derivative of this count rate is calculated as shown in Fig.~\ref{fig:coherent_state_characterisation}~(c). This represents the number of events which were additionally rejected at a given delay step and should thus mirror the distribution of rising edge timings generated by the SNSPD. When comparing this derivative to the rising-edge timing distribution obtained from a separate TDC measurement, it can be seen that the structure and distance of the individual photon-number peaks is well preserved by the selection circuit. There is however a slight degradation in jitter performance leading to blurring between the peaks for two and three photons. This is likely caused by the gray zone of the NAND gate. Nonetheless, the single- and two-photon peaks remain well separated.

For the subsequent demonstration, we will use the circuit to select single-photon events exclusively. Therefore, we always set the delay of the laser clock to the minimum between the single- and two-photon peaks, as indicated in Fig.~\ref{fig:coherent_state_characterisation}. 

To tune the precise timing cutoff of our circuit it is always necessary to adjust the timing of the laser clock relative to the SNSPD signal. For the characterization presented here we employ a high-precision delay generator outside the cryostat to scan the laser clock delay. While this significantly increases the complexity of the readout system, it is required to ensure comparability with the results obtained by the TDC. The delays can however also be tuned in the cryostat by varying the supply voltage of selective logic elements \cite{Hummel2023}, as shown in appendix~\ref{sec:delay_tuning}. This in situ delay tuning by supply voltage can be used when trying to analyze multiple SNSPD signals in parallel to increase scalability.

We also characterize the power dissipated by the circuit by measuring the supply voltages and currents entering the cryostat. This yields a total power dissipation of \SI{12.8}{mW}, with \SI{3.1}{mW} for the amplifier, \SI{4.8}{mW} for the trigger and logic circuits and \SI{4.9}{mW} for the signal transmission outside the cryostat. This is compatible with the \SI{200}{mW} cooling power of the \SI{4}{K} cooling stage to which the circuit is mounted.

In the context of signal processing latency for applications like feedforward, we also characterize the delay between the photon detection at the SNSPD and the output signal of the photon-number filter. For this purpose, we first calibrate the delay between the laser trigger signal and the signal recorded at the oscilloscope by substituting the SNSPD and filter circuit with a photodiode (Thorlabs DET08CFC)(5GHz). This sets our calibration plane at the fiber and coaxial connectors of the cryostat. We then measure the delay of the fiber/coaxial cables into the cryostat using reflection measurements to move the calibration plane to the SNSPD and PCB-connector. With this calibration we can finally measure the delay of the photon filter relative to the laser trigger signal and compute the latency. We measure a latency of \SI{13.9\pm0.3}{ns}, which is comparable to similar cryogenic feedforward circuits, which report electrical delays of \SI{20\pm2}{ns} \cite{Thiele2025}.

\section{application as a single photon herald}

\begin{figure*}[ht]
    \centering
    \includegraphics[width=1\linewidth]{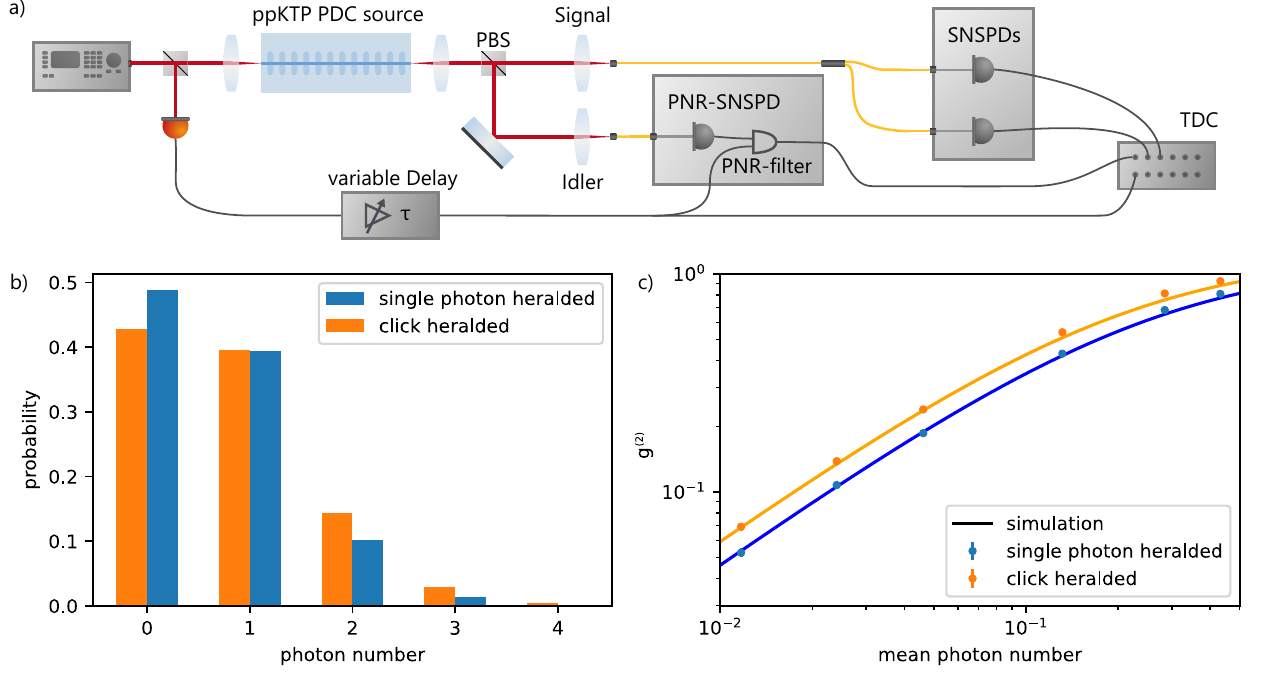}
    \caption{a) The setup used to test the single photon heralding. The type-II SPDC source creates a photon-number correlated state between two polarization modes, which is split by the polarizing beamsplitter (PBS). Measurement of the photon-number in the idler arm provides photon-number information of the signal arm, allowing for single photon heralding. b) The heralded photon-number distribution of the SPDC signal arm when heralding on either single photons or all detection events. The data shown here is for the highest pump power setting with a mean detected photon number of ~0.45. c) The normalized second order correlation function $g^{(2)}$ measured at different SPDC brightness settings. The x-axis shows the unheralded mean photon number measured in the signal arm.}
    \label{fig:g2_plot}
\end{figure*}

For this experiment, we use a type-II SPDC source and separate the signal and idler modes with a polarizing beamsplitter. The idler mode is routed to the photon-number-selective detector system, while the photon number in the signal mode is measured with two commercial photon-number-resolving SNSPDs, whose photon-number information is analyzed in software. The single-photon-selective signal from our circuit is then used to herald the presence of a single photon. In addition, the timing filter can also be configured to herald every detection event by shifting the timing cutoff to earlier times, as shown in Fig.~\ref{fig:coherent_state_characterisation} for \SI{0}{ps} laser clock delay.

We vary the pump power of the SPDC process to access different mean photon numbers created in the source, and thereby access larger contributions from higher-order photon-number generation events. This affects the heralded single-photon purity as there is a trade-off between single-photon purity and source brightness, which single-photon-specific heralding can alleviate.

The timing cutoff calibration for single-photon selection is performed as described in Sec.~\ref{ch:characterisation}. This calibration must be performed for all mean photon numbers as the timing of the SNSPD pulses can shift up to hundreds of picoseconds with increased detection rate. This is likely due to the charging of AC-coupling capacitors which shifts the baseline voltage of the SNSPD signal and thereby alters the effective trigger level, as has been observed in previous experiments \cite{Schapeler2026}. This could potentially be alleviated by using a fully DC-coupled amplifier chain. This is, however, outside the scope of this experiment.

With this setup, we can now measure the changes in the heralded photon-number distribution caused by single-photon-selective heralding. For this one can either compare the photon-number statistics directly, as shown in Fig.~\ref{fig:g2_plot}~(b), or compute the second order correlation function $g^{(2)}$, shown in Fig.~\ref{fig:g2_plot}~(c), which serves as an indicator for the single-photon purity of the SPDC source.

For the heralded photon-number statistics it can be seen that operating the circuit to selectively pass only single-photon events reduces the multi-photon to single-photon ratio by up to 34\% while having an almost identical single-photon heralding probability. Likewise for the second order correlation function $g^{(2)}$ it can be seen that heralding on single photons reduces $g^{(2)}$ by a factor of \SI{24\pm5}{\%} for the same optical power pumping the PDC process. The observed improvement is not primarily limited by the photon-number resolution of the detector but instead by the losses prior to detection of about 64\%. These losses occur primarily due to the coupling efficiency of the idler beam into fiber and cause multi-photon events to be misidentified as single photons due to some photons being lost prior to detection. 

Our improvement in $g^{(2)}$ is favorably comparable to that reported for other systems utilizing intrinsic photon-number resolution with postprocessing at \SI{13\pm0.4}{\%} \cite{Sempere-Llagostera2022}, \SI{20}{\%} \cite{Davis2022} and the improvement shown with multi-pixel detectors at \SI{26.9\pm0.1}{\%} \cite{Stasi2023}. This demonstrates the capability of the real-time readout system to achieve state-of-the-art photon-number selection while providing the advantage of low latency within the cryostat. This enables integration of this system in more complex feedforward systems like a single-photon-selective SPDC source multiplexer \cite{Kaneda2019}.

\section{Discussion}

In this work, we have demonstrated an electrical readout scheme capable of \emph{in situ} photon-number discrimination from an intrinsically photon-number-resolving SNSPD. We investigate the photon-number resolution of the readout system and found good agreement with a separate analysis performed with a TDC. Furthermore, we demonstrate the possibility to utilize this readout to herald single photons from an SPDC source.

Currently, the system can only discriminate between the presence of n=0,1 and 2+ photons and cannot distinguish between n=2 and n=3+ photons, as would be required in some applications \cite{DeGliniasty2024}. Extending the system, however, is straightforward, as adding an identical copy of the discrimination circuit with a delay set to discriminate between the 2-photon and 3-photon peaks would readily provide this capability.

The cryogenic readout system shown here enables latency-critical applications, such as feedforward systems, that utilize the intrinsic photon number resolution of SNSPDs. Realistic applications, however, will require a multitude of PNR detectors and corresponding readout systems, requiring significant scale up. The scalability of this approach can be enhanced by integrating the time discrimination system on a single silicon die. This integration can improve the packaging density as well as the resolution of the system. For example, components such as the NAND gate could be optimized on a transistor level or even replaced with other logic elements, such as a D-latch, depending on the requirements of the feedforward system.

In order to build scalable systems with intrinsic photon-number resolution, the power consumption of the analysis circuit must also be significantly reduced. Other studies have already demonstrated cryo-CMOS systems for SNSPD readout with a power dissipation as low as \SI{33}{\mu W} in an integrated platform \cite{Carboni2025}. Amplifiers based on superconducting nTrons have also shown great success for SNSPD readout, enabling a very low power dissipation of \SI{e-17}{J/bit} \cite{McCaughan2014}. Applying these novel technologies presents significant opportunities to engineer low-power systems based on the photon-number readout method presented in this paper.

This PNR readout circuit can also find application beyond feedforward systems. For general-purpose, non-latency-critical detector readout, the filtering circuit can provide a digitally encoded photon-number signal. Such signals have a reduced bandwidth and noise requirement compared to the raw analog signal generated by the SNSPD. Additionally, multiple such signals can be transmitted over a single cable using additional data serialization circuits. This can be utilized in scaling up PNR detector systems based on intrinsic photon-number resolution.

\section{Data Availability}
The data that support the findings of this study are openly available in Zenodo at https://doi.org/10.5281/zenodo.24471773.

\begin{acknowledgments}
The authors thank Frederik Thiele and Maximilian Protte for helpful discussions. This work was partially funded by the European Union [European Research Council (ERC), “Quantum Engineering of Superconducting Array Detectors In Low-Light Applications” (QuESADILLA), Grant No. 101042399]. The views and opinions expressed are, however, those of the author(s) only and do not necessarily reflect those of the European Union or the European Research Council Executive Agency. Neither the European Union nor the granting authority can be held responsible for them. This work was supported by the German Federal Ministry of Research, Technology and Space (BMFTR) through TUF-ToPiQC (13N17238). NIST work was funded entirely by the US government. The use of trade names is intended to allow the measurements to be appropriately interpreted and does not imply endorsement by the US government, nor does it imply these are necessarily the best available for the purpose used here.
\end{acknowledgments}

\appendix

\section{alternate delay tuning}\label{sec:delay_tuning}

As an alternative to tuning the delay of the laser timing reference with an external delay generator, the delay of the reference can also be tuned by varying the supply voltage of some circuit elements. For larger versions of our photon-number discriminator with multiple channels and higher photon-number resolution, this presents a more scalable solution compared to using an external delay generator for every channel. 

This delay tuning has been tested in the same manner as the delay tuning with the external delay generator. The detector is again illuminated by a pulsed attenuated coherent state. The delay generator, however, only acts to generate a coarse delay to compensate for different path lengths in the optical and electrical pulses. The fine tuning of the delay is instead performed by using the $V_{FB}$ supply voltage of the trigger. This changes the delay generated by the trigger circuit and therefore delays the incoming SNSPD pulse.

\begin{figure}[ht]
    \centering
    \includegraphics[width=1\linewidth]{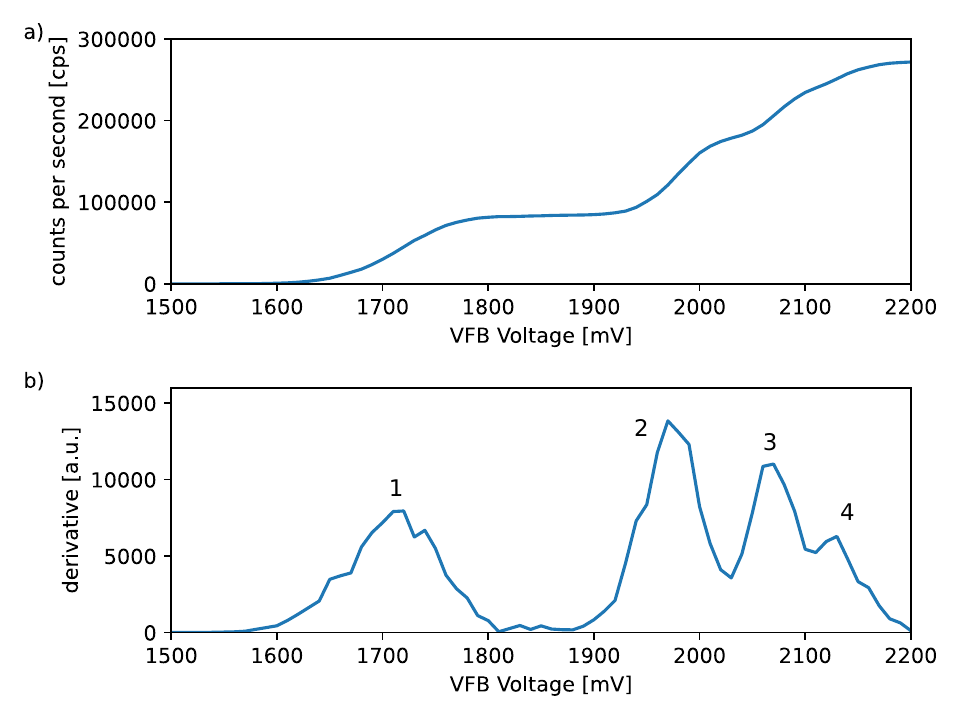}
    \caption{a) The count rate of the SNSPD after the photon discrimination circuit when varying the trigger supply voltage VFB. b) The derivative of the count rate shown in a). Clear peaks corresponding to photon numbers can be observed.}
    \label{fig:delayscan}
\end{figure}

The resulting count rate of this tuning approach can be seen in Fig. \ref{fig:delayscan}. The derivative of the count rate exhibits the same shape as when using the external delay generator. This shows the feasibility of using the supply voltage for delay tuning instead of the external delay generator.

\section{$g^{(2)}$ calculation and simulation}

For the measurement of the heralded $g^{(2)}$, we employ a standard Hanbury Brown-Twiss (HBT) interferometer architecture. The SNSPDs used here were, however, treated as PNR detectors. This allows us to directly measure the photon-number statistics of the heralded single photons. 

From this photon-number distribution $P(n)$ we can now calculate the $g^{(2)}$ of the state according to 
\begin{equation}
    g^{(2)}=\frac{\sum^{\infty}_{n=0}n(n-1)P(n)}{(\sum^{\infty}_{n=0}nP(n))^2}
\end{equation}
\cite{Migdall2013}. Since the detectors used here are not perfect PNR detectors, we also correct the photon-number distribution $P(n)$ by the POVM matrix of the detector. This POVM can be theoretically constructed from the convolution of the POVMs of two ideal PNR detectors with a resolution of 2 photons \cite{Davis2025}.

For the simulation of the $g^{(2)}$ values we use the same method as Thiele et al. \cite{Thiele2025}. We, however, assume a perfect identity matrix for the detector POVM, as there is negligible overlap between the two-photon and single-photon timing peaks. The losses acting on the idler and signal modes are based on the measured Klyshko efficiency of the setup. The statistics of the SPDC source are modeled as three equal strength Schmidt modes, to match the observed unheralded $g^{(2)}$ of ~1.3. The assumption of equal squeezing in all three Schmidt modes is made because we do not have a full Schmidt decomposition available for the source.

\section{detector characterization}

\begin{figure}[ht]
    \centering
    \includegraphics[width=0.7\linewidth]{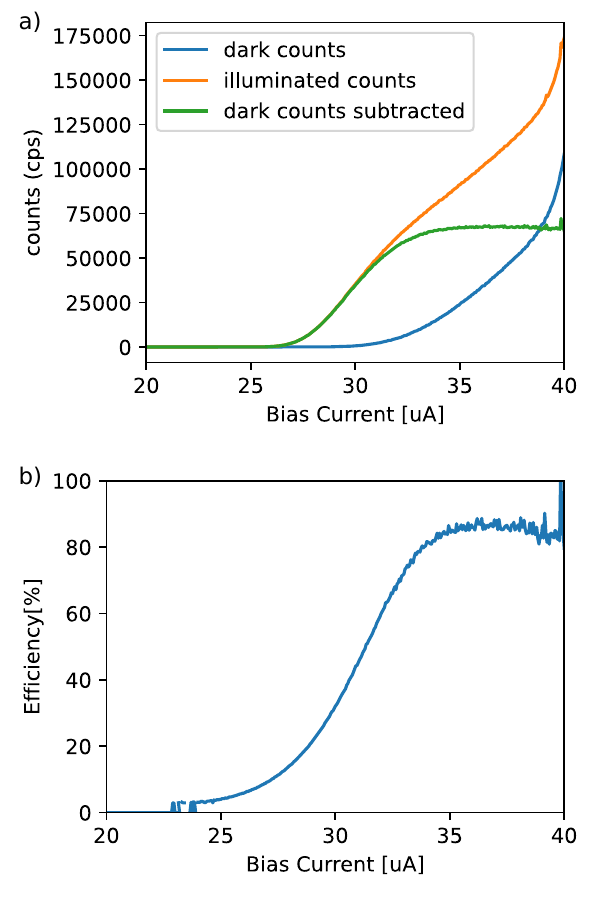}
    \caption{a) The count rate of the detector measured against its supplied bias current. The countrate starts to plateau around \SI{35}{\mu A}, indicating a saturated detection efficiency. b) The calibrated efficiency of the detector.}
    \label{fig:det_char}
\end{figure}

Prior to the experiment, the SNSPD was characterized for efficiency  and dark counts. Those measurements are shown in Fig. \ref{fig:det_char}. For the efficiency measurement, the detector was illuminated by an attenuated pulsed laser. The mean photon number per pulse of the laser was calibrated with a different SNSPD from Single Quantum, for which a factory calibrated efficiency is available. The high dark count rate of the detector is not detrimental in this experiment, as the short timing window around the laser clock eliminates nearly all dark counts. 

\section{circuit schematic and operating parameters}

\begin{figure}[ht]
    \centering
    \includegraphics[width=1\linewidth]{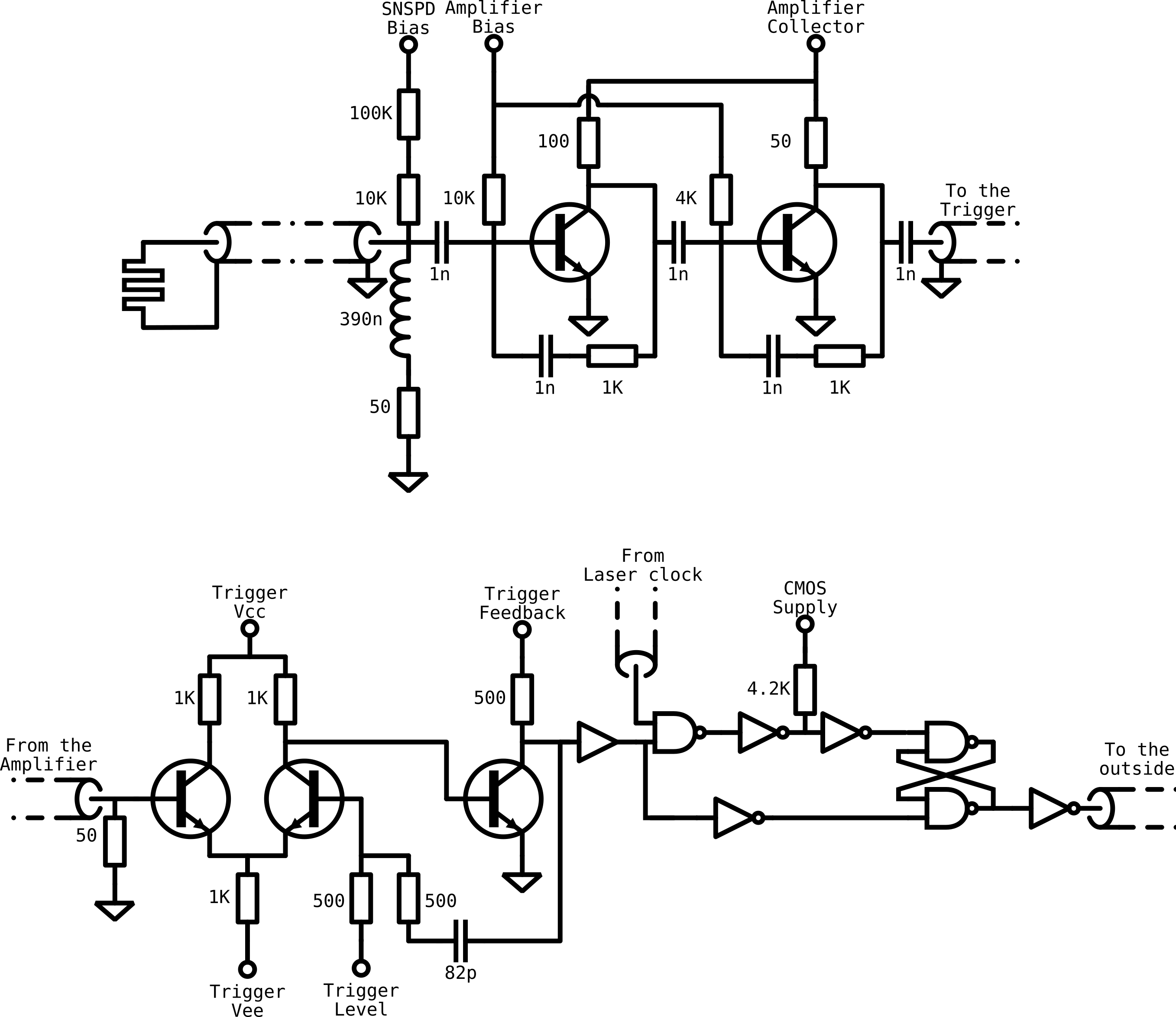}
    \caption{The schematic of the circuit used to perform the photon-number selection. AC-Bypass capacitors are omitted for readability.}
    \label{fig:circuit}
\end{figure}

A full schematic of the circuit used to perform the photon-number selection is shown in Fig.~\ref{fig:circuit}. The design of the trigger is adapted from a previous work \cite{Thiele2025}. To ensure compatibility with the cryogenic environment, we use NiCr thin film resistors, NP0 ceramic capacitors and air core inductors for the passive components. For the active components we use BFP 640 Heterojunction Bipolar Transistors (HBT) and CMOS logic gates from the Onsemi TinyLogic ULP-A series.

For calculating the power dissipation of the circuit we can use the values given in Table~\ref{tab:parameters} and use $P=U\cdot I$ to calculate the dissipated power of the circuit (assuming no ohmic loss in the cables). We add one additional correction to the power budget as the digital signals transmitted outside the cryostat partially dissipate over the \SI{50}{\Omega} termination at the oscilloscope/TDC. Based on the \SI{200}{ns} pulse width \SI{100}{kHz} repetition rate and \SI{2.5}{V} amplitude, this power dissipation can be calculated to be \SI{2.5}{mW}.

\begin{table}[h]
    \centering
    \begin{tabular}{ccc}
        Electrical Connection & Voltage [V] & Current [mA] \\
        Amplifier Bias & $1.250\pm0.001$ & $0\pm1$ \\
        Amplifier Collector & $0.580\pm0.001$ & $5\pm1$  \\
        SNSPD Bias & $4.000\pm0.001$ & $0.042\pm0.005$ \\
        Trigger Level & $0.065\pm0.001$ & $0.014\pm0.005$ \\
        Trigger Feedback & $2.500\pm0.001$ & $0.162\pm0.005$ \\
        Trigger Vcc & $1.400\pm0.001$ & $0.411\pm0.005$ \\
        Trigger Vee & $1.400\pm0.001$ & $0.391\pm0.005$ \\
        CMOS supply & $3.300\pm0.001$ & $3\pm1$ \\
        Laser clock & $2.000\pm0.001$ & 2V/50$\Omega$ (@ 1\% duty cycle) \\
    \end{tabular}
    \caption{The electrical supply voltages and currents used to operate the circuit.}
    \label{tab:parameters}
\end{table}

\section{circuit extension for more photon number resolution}

\begin{figure}[ht]
    \centering
    \includegraphics[width=1\linewidth]{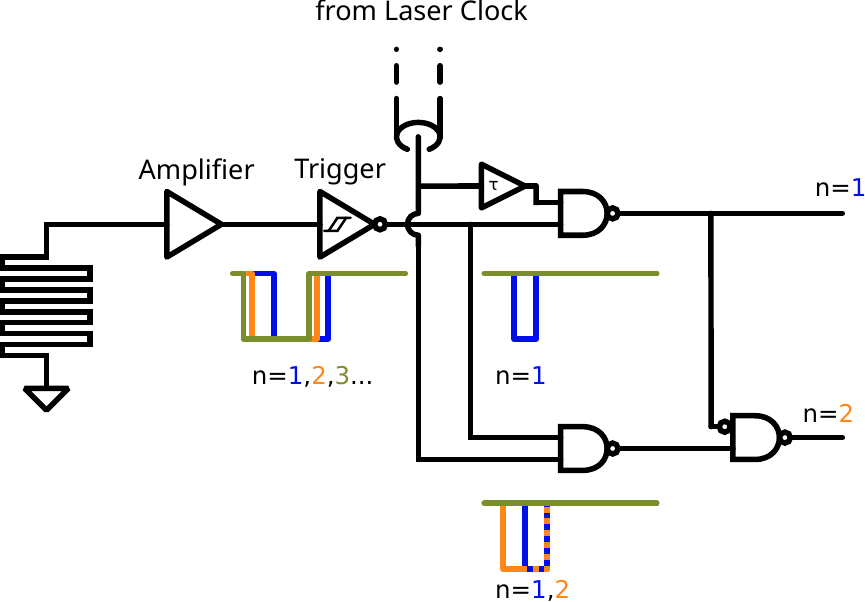}
    \caption{The schematic of the circuit extended to provide more photon number resolution.}
    \label{fig:circuit_ext}
\end{figure}

In order to decode more photon numbers from the timing signal of the SNSPD, an additional copy of the circuit discussed in the main text can be added. A schematic of this is shown in Fig.~\ref{fig:circuit_ext}. The capability to resolve two photon events is provided by an additional NAND gate. This gate is supplied with a clock signal, which arrives before the clock for single photon events and thereby selects events corresponding to n=1 and n=2 photons. Using another NAND gate with an inverted input the n=2 signal can be isolated. 

\bibliography{bibliography.bib}

\end{document}